\documentclass[12pt,a4paper]{article}
\usepackage{amssymb,amsmath}
\usepackage{a4wide}
\usepackage{epsfig}
\usepackage{graphics}
\usepackage{subfigure}
\usepackage{amsfonts}

\title{Dynamics of Periodic Monopoles}
 \author{Derek Harland${}^a$\footnote{Corresponding author.
   Emails: harland@math.uni-hannover.de, richard.ward@durham.ac.uk}\quad
     and R.\ S.\ Ward${}^b$\\
   ${}^a$ {\small Institut f\"ur Theoretische Physik, Leibniz Universit\"at Hannover, 30167 Hannover, Germany}\\
   ${}^b$ {\small Dept of Mathematical Sciences, University of Durham,
        Durham DH1 3LE, UK}
}
\date{22nd April 2009}

\newcommand{\RR}{{\mathbb{R}}}
\newcommand{\CC}{{\mathbb{C}}}
\newcommand{\pa}{\partial}
\newcommand{\ii}{{\rm i}}
\newcommand{\ee}{{\rm e}}
\newcommand{\half}{{\textstyle\frac{1}{2}}}
\newcommand{\diag}{\mathop{\rm diag}\nolimits}
\newcommand{\tr}{\mathop{\rm tr}\nolimits}
\renewcommand{\Re}{\mathop{\rm Re}\nolimits}
\renewcommand{\Im}{\mathop{\rm Im}\nolimits}
\begin{document}

\maketitle
\abstract{
\noindent BPS monopoles which are periodic in one of the spatial
directions correspond, via a generalized Nahm transform, to solutions
of the Hitchin equations on a cylinder. A one-parameter family of
solutions of these equations, representing a geodesic in the
2-monopole moduli space, is constructed numerically. It corresponds
to a slow-motion dynamical evolution, in which two parallel monopole
chains collide and scatter at right angles.
}
\vskip 1truein
\noindent Keywords: monopoles, moduli-space.
\vskip 1truein
\noindent PACS 11.27.+d, 11.10.Lm, 11.15.-q

\newpage

\section{Introduction}

The idea of understanding the slow-motion dynamics of BPS monopoles by
studying the geodesics on the moduli space of static monopoles
has been extensively exploited for monopoles in $\RR^3$
--- see \cite{MS04} for a review. For example, much is known
about the 4-dimensional moduli space of centred 2-monopoles in $\RR^3$
(the Atiyah-Hitchin space), about its geodesics, and hence about
the mutual scattering of two monopoles.
The purpose of this note is to look at an example of the corresponding
thing for periodic monopoles, {\sl ie.}\ monopoles on $\RR^2\times S^1$.
(This could be visualized as the transverse scattering of two parallel
chains of BPS monopoles.)
In this case \cite{CK02}, the moduli space of centred 2-monopoles is again
4-dimensional, an ``ALG gravitational instanton of type $D_0$'',
but much less is known about its metric;
in particular, the geometry is believed to have no continuous symmetries
(unlike for the Atiyah-Hitchin metric). But it does have discrete isometries,
and this enables us to identify some geodesic submanifolds and geodesics.
We shall see that one such geodesic corresponds to the head-on collision
of two monopoles, resulting (as usual) in $90^{\circ}$ scattering.

The approach is to use a generalized Nahm transform \cite{CK01},
which says that periodic monopoles correspond to certain solutions
of the Hitchin equations. The latter are what one gets by reducing
the (4-dimensional) self-dual Yang-Mills equations down to two dimensions.
This system has been studied for some time. Long ago,
it was noted there there are no finite-energy solutions of the SU(2)
Hitchin equations on $\RR^2$ \cite{L77, S84}.
Hitchin's comprehensive paper, investigating the equations on compact
Riemann surfaces, appeared in 1987 \cite{H87}.
For non-compact gauge groups such as SO(2,1), it was pointed out more
recently that finite-energy solutions on $\RR^2$ do exist \cite{MJ06}.

In our case \cite{CK01}, we are interested in the SU(2) Hitchin
equations on the cylinder $\RR\times S^1$. These equations are analysed
in section 2 below; section 3 reviews how their solutions may be used
to construct periodic monopoles, and defines a natural totally-geodesic
surface ${\cal S}$ in the moduli space; and section 4 describes
the monopoles corresponding to points of ${\cal S}$, and in particular
a geodesic in ${\cal S}$ which represents the head-on collision of
two periodic monopoles. Finally, in section 5, we show how a particular explicit
solution of the Hitchin equations may be obtained as the $N\to\infty$ limit
of the Nahm data for a finite chain of $N$ monopoles.

Except for this special case, the Hitchin-equation solutions are not
explicit, and have to be obtained numerically. The Nahm transform
which generates the monopole fields is also implemented numerically,
to produce the picture of monopole scattering. This is analogous to
what was done to obtain examples of $N$-monopole scattering in $\RR^3$
--- see \cite{MS04} for examples.

%%%%%%%%%%%%%%%%%%%%%%%%%%%%%%%%%%%%%%%%%%%%%%%%%%%%%%%%%%%%%%%%%%%%

\section{The SU(2) Hitchin Equations}

The Hitchin equations may be viewed as a 2-dimensional reduction of the
4-dimensional self-dual Yang-Mills equations $F_{12}=F_{34}$,
$F_{13}=F_{42}$, $F_{14}=F_{23}$. Assume that the gauge potential $A_\mu$
depends only on the two coordinates $(x^1,x^2)$, and write $s=x^1+\ii x^2$ and
$\Phi=A_3-\ii A_4$.  Then the SDYM equations reduce to the Hitchin equations
\begin{eqnarray}  %\label{covariant derivative}
  D_{\bar s}\Phi &=& 0, \label{H1} \\
  F_{12} &=& \half\ii [\Phi,\Phi^*]. \label{H2}
\end{eqnarray}
Here $D_{\bar s}\Phi:=\pa_{\bar s}\Phi+[A_{\bar s},\Phi]$, and $\Phi^*$ denotes
the complex conjugate transpose of $\Phi$. Take the gauge group to be
SU(2); the fields $\Phi$ and $A_{\bar s}$ are trace-free $2\times2$ complex
matrices, and are smooth on some surface $\Sigma$ having local coordinates
$(s,{\bar s})$.

We may partially solve the Hitchin equations as follows.
First, note that (\ref{H1}) implies that $\det(\Phi)$ is holomorphic, so
we write $\det(\Phi)=-H(s)$, and think of the holomorphic function $H(s)$
as being fixed.
Next, we can use the gauge freedom to diagonalize the gauge field
$F_{12}=-2\ii F_{s{\bar s}}$ on $\Sigma$ (as long as there is no global
topological obstruction --- but for the case we are interested in, namely
$\Sigma=\RR\times S^1$, there is no such obstruction).
So we write $F_{12}=\ii B \sigma_3$, where $\sigma_3=\diag(1,-1)$, and
where $B$ is a real-valued function on $\Sigma$. Let us assume that
$B$ is not identically zero (the case where $B=0$ is easy to solve locally).
Then the residual gauge freedom is a
local O(2), consisting of a local U(1)
\begin{equation}  \label{local U1}
  \Phi\mapsto U^{-1}\Phi U,\quad\quad U:=\exp[\ii\, u(s,{\bar s})\,\sigma_3],
\end{equation}
plus a reflection. The second Hitchin equation (\ref{H2}) now implies
that $\Phi$ has the form
\[
  \Phi=\left[\begin{array}{cc}
         0 & f \\
         g & 0 \end{array}\right],
\]
where $fg=H$. The residual gauge freedom acts on the functions $f$ and $g$
by
\begin{eqnarray}
f &\mapsto& f\ee^{-2\ii u},\quad g\mapsto g\ee^{2\ii u}\quad{\rm (local\ U(1))},
     \label{gaugeu1}\\
f &\mapsto& g,\quad g\mapsto f\quad{\rm (reflection)}\label{gaugerefl}.
\end{eqnarray}

The first Hitchin equation (\ref{H1}) is then equivalent to the gauge
potential $A$ having the form $A_{\bar s}=a\sigma_3 + \alpha\Phi$, where
$a(s,{\bar s})$ and $\alpha(s,{\bar s})$ are complex-valued functions, and
where $a$ given by
\begin{equation}  \label{afg eqn}
   2a = g^{-1}\pa_{\bar s}g = -f^{-1}\pa_{\bar s}f.
\end{equation}
Finally, what remains of (\ref{H2}) is equivalent to the two equations
\begin{equation}  \label{DeltaEqn1}
   \Delta\left[\Re\log(g/f)\right] = 2(1+4|\alpha|^2)(|g|^2-|f|^2),
\end{equation}
where $\Delta=\pa_1^2+\pa_2^2=4\,\pa_s\,\pa_{\bar s}$ is the 2-dimensional
Laplacian; and
\begin{equation}  \label{AlphaEqn1}
   {\bar f}^{-1}\pa_s(|f|^2\alpha)+g^{-1}\pa_{\bar s}(|g|^2{\bar\alpha})=0.
\end{equation}
From (\ref{afg eqn}) it follows that, in order for $a$ to be
smooth, the functions $f$ and $g$ must have the form
\begin{equation}  \label{fg1}
  f=\mu_+\ee^{\psi/2}, \quad g=\mu_-\ee^{-\psi/2},
\end{equation}
where $\mu_\pm$ are holomorphic functions with $\mu_+\mu_-=H$, and
$\psi=\psi(s,{\bar s})$ is a smooth function. Using the residual gauge
freedom (\ref{gaugeu1}), we can set $\psi$ to be real-valued. Then
(\ref{DeltaEqn1}) becomes
\begin{equation}  \label{psi eqn}
   \Delta\psi = 2(1+4|\alpha|^2)(\xi_+\ee^\psi - \xi_-\ee^{-\psi}),
\end{equation}
where $\xi_\pm = |\mu_\pm|^2$; and (\ref{AlphaEqn1}) can be re-written as
\begin{equation}  \label{AlphaEqn2}
   \ee^{-\psi/2}\pa_s(\ee^{\psi}\mu_+\alpha)+
     \ee^{\psi/2}\pa_{\bar s}(\ee^{-\psi}{\bar\mu_-}{\bar\alpha})=0.
\end{equation}

To summarize: given $H(s)$, we first choose a holomorphic splitting
$H=\mu_+\mu_-$ (essentially, the only choice is which zeros of $H$
to include in $\mu_+$, and which to include in $\mu_-$). Then
the Hitchin equations (\ref{H1}, \ref{H2}) are equivalent to the coupled
system (\ref{psi eqn}, \ref{AlphaEqn2}) for the real function $\psi$ and
the complex function $\alpha$. Finally, the functions
$f,g,a$ (and hence $\Phi$ and $A_{\bar s}$) are
determined by (\ref{fg1}) and $4a=-\pa_{\bar s}\psi$. Note that
the functions $\psi$ and $\alpha$ are gauge-invariant, except that $\psi$
changes sign under the reflection (\ref{gaugerefl}).

%%%%%%%%%%%%%%%%%%%%%%%%%%%%%%%%%%%%%%%%%%%%%%%%%%%%%%%%%%%%%%

\section{Nahm Transform}

In this section, we review how certain solutions of the SU(2) Hitchin
equations correspond, via a generalized Nahm transform, to
centred periodic monopoles of charge 2. Such a periodic monopole solution
consists of an SU(2) gauge field $\hat{A}_j=\hat{A}_j(x,y,z)$, and
a Higgs field $\hat{\Phi}=\hat{\Phi}(x,y,z)$ in the adjoint representation,
satisfying
\begin{itemize}
\item $\hat D_j\hat{\Phi}=-\half\varepsilon_{jkl}\hat{F}_{kl}$, where
   $\hat{F}_{jk}$ is the gauge field obtained from $\hat{A}_j$;
\item $\hat{\Phi}$ and $\hat{A}_j$ are smooth, and periodic in
   $z$ with period $2\pi$;
\item locally in some gauge, $\hat{\Phi}$
   and $\hat{A}_j$ satisfy the boundary conditions
   $\hat{\Phi}-(\ii/\pi)(\log\rho)\sigma_3\to0$,
   $\hat{A}_x\to0$, $\hat{A}_y\to0$ and
   $\hat{A}_z-(\ii\theta/\pi)\sigma_3\to0$ as $\rho\to\infty$.
   (Here $\rho$ and $\theta$ are polar coordinates:
   $x+\ii y=\rho\ee^{\ii\theta}$.)
\end{itemize}

A generalized Nahm transform \cite{CK01} relates such monopoles to
certain solutions of the SU(2) Hitchin equations on the cylinder
$\RR\times S^1$. Let $r\in\RR$, and $t$ with period 1,
denote the coordinates on this cylinder. The solutions are
required to satisfy
\begin{itemize}
  \item $H(s)=2\cosh(2\pi s)-K$, for some constant $K\in\CC$; and
  \item $F_{12}\to0$ as $r\to\pm\infty$.
\end{itemize}
In the first of these conditions, we could introduce a couple of
parameters, by writing $H(s)=2C\cosh(2\pi s)-K$, where $C$ is a complex
constant. Then $|C|$ would determine the ratio between the monopole size
and the spatial period, while $\arg(C)$ would determine a spatial orientation.
In the monopole picture, these two real parameters would show up by
a slight alteration in the form of the boundary conditions on $\hat{\Phi}$
and $\hat{A}_j$.
For simplicity, both of the parameters are omitted here, but it
would be straightforward to re-introduce them. It is also worth pointing
out that restricting to SU(2) rather than U(2) Hitchin fields has the
effect of centring the corresponding monopoles, {\sl ie.}\ removing the
translation freedom in $x,y,z$.

The details of the transform by which one obtains $\hat{\Phi}$ and
$\hat{A}_j$ from $\Phi$ and $A_{\bar s}$ are as follows. Write
$\zeta=x+\ii y$. Let $\Psi_+$
and $\Psi_-$ be $2\times2$ matrices satisfying the linear system 
\begin{eqnarray}
 2\pa_{\bar s}\Psi_+ + 2A_{\bar s}\Psi_+ - z\Psi_+ + \zeta\Psi_-
   - \Phi\Psi_- &=& 0,  \label{PsiEqn+}\\
 2\pa_{s}\Psi_- + 2A_{s}\Psi_- + z\Psi_- + {\bar\zeta}\Psi_+
   - \Phi^*\Psi_+ &=& 0,  \label{PsiEqn-}
\end{eqnarray}
as well as the normalization condition
$\langle\Psi_+,\Psi_+\rangle+\langle\Psi_-,\Psi_-\rangle={\bf I}$.
Here {\bf I} is the $2\times2$ identity matrix, and
$\langle\, ,\,\rangle$ is the $L^2$ inner product
$\langle\Theta,\Gamma\rangle =
     \int_{-\infty}^{\infty}\int_0^1 \Theta^*\, \Gamma \,dt\,dr.$
Let $\Psi$ be the $4\times2$ matrix obtained by adjoining $\Psi_+$
and $\Psi_-$, in other words
\[
  \Psi=\left[\begin{array}{c}\Psi_+\\\Psi_-\end{array}\right].
\]
Then $\hat{\Phi}$ and $\hat{A}_j$ are given by
\begin{equation}  \label{MonFields}
  \hat{\Phi}=\ii\langle\Psi,r\Psi\rangle,\quad
    \hat{A}_j=\langle\Psi,\pa_j\Psi\rangle.
\end{equation}

The moduli space $\widehat{\cal M}$ of centred periodic 2-monopoles
is 4-dimensional, and it has a natural hyperk\"ahler metric \cite{CK02}.
This space is the same (via the Nahm transform) as the moduli space
${\cal M}$ of solutions of the Hitchin equations satisfying the
various conditions listed above. (Strictly speaking, it has not been
proved that the natural hyperk\"ahler metrics on $\widehat{\cal M}$
and ${\cal M}$ are isometric, but this is expected, on general grounds,
to be true.) Two of the moduli consist
of the real and imaginary parts of $K$, and the remaining two
determine the function $\alpha$.

In what follows, we shall concentrate on the surface ${\cal S}$ in
$\widehat{\cal M}\cong{\cal M}$ corresponding to $\alpha=0$.
This surface ${\cal S}$ is a totally-geodesic submanifold of
${\cal M}$: to see this, we may reason as follows.
Note, first, that rotation of the monopole field by $\pi$ about the
$z$-axis ({\sl ie.}\ $\zeta\mapsto-\zeta$ or $\theta\mapsto\theta+\pi$)
preserves the boundary condition on $(\hat{\Phi},\hat{A}_j)$;
the change in asymptotic behaviour can be compensated by a gauge
transformation. So this map is an isometry of $\widehat{\cal M}$.
From (\ref{PsiEqn+}, \ref{PsiEqn-}) we see that it corresponds
to $\Psi_+\mapsto\Psi_+$ and $\Psi_-\mapsto-\Psi_-$, together with the
map
\begin{equation}  \label{alpha to minus alpha}
  \Phi\mapsto-\Phi, \quad A_{\bar s}\mapsto A_{\bar s}
 \qquad \Longleftrightarrow \qquad \psi\mapsto\psi, \quad \alpha\mapsto-\alpha.
\end{equation}
So (\ref{alpha to minus alpha}), which clearly preserves the Hitchin
equations and their boundary conditions, is also an isometry of the
moduli space ${\cal M}$. It follows that the surface ${\cal S}$ corresponding
to $\alpha=0$ is a totally-geodesic submanifold of ${\cal M}$;
or equivalently that the surface corresponding to monopoles
invariant under rotations by $\pi$ about the periodic axis, is a
totally-geodesic submanifold of $\widehat{\cal M}$.

%%%%%%%%%%%%%%%%%%%%%%%%%%%%%%%%%%%%%%%%%%%%%%%%%%%%%%%%%%%%%%%%%%

\section{Monopoles and Scattering}

For our subfamily ${\cal S}$ of solutions with $\alpha=0$, the Hitchin
equations reduce, in effect, to
\begin{equation}  \label{PsiEqn2}
   \Delta\psi = 2(\xi_+\ee^\psi - \xi_-\ee^{-\psi}).
\end{equation}
In this section, we examine solutions of (\ref{PsiEqn2}), and describe the
corresponding periodic monopoles; in particular, a geodesic in ${\cal S}$
representing a head-on scattering process.

First, there is the matter of defining $\mu_+(s)$ and $\mu_-(s)$, which
effectively means choosing their zeros. The function $H(s)$ has two zeros,
and we shall adopt the most obvious choice of allocating one of them
to $\mu_+(s)$ and the other to $\mu_-(s)$: set
$\mu_{\pm} = \ee^{\pi s} - \lambda_{\pm} \ee^{-\pi s}$,
where $\lambda_{\pm}=(K\pm\sqrt{K^2-4})/2$.
In the square root, take $\sqrt{K^2-4}$ to be continuous on the complement
of the cut $K\in(-2,2)$, with $\Im(\sqrt{K^2-4})\geq0$ for $\Im(K)\geq0$.

The boundary condition $F_{12}\to0$ as $r\to\pm\infty$
is equivalent to a boundary condition on $\psi$, namely
$\xi_+\ee^\psi - \xi_-\ee^{-\psi} \to0$ as $r\to\pm\infty$;
with our choice of $\mu_{\pm}(s)$ this gives
$\psi\to0$ as $r\to+\infty$ and $\psi\to\log|\lambda_-/\lambda_+|$
as $r\to-\infty$.
It is easy to see that, with this boundary condition (and fixing $K$),
any small perturbation $\psi\mapsto\psi+\delta\psi$ of a solution $\psi$
of (\ref{PsiEqn2}) has to be trivial.
For if $\delta\psi$
is such a perturbation, with $\delta\psi\to0$ as $r\to\pm\infty$,
then $L(\delta\psi)=0$, where $L$ is the strictly-negative operator
$L=\Delta-2(\xi_+\ee^\psi + \xi_-\ee^{-\psi})$; so the only solution
is $\delta\psi=0$. 

Given that $\psi(r,t)$ is to be periodic in $t$, we need to change gauge in
order for $f$ and $g$ to be periodic: instead of the expressions (\ref{fg1}),
we set
\begin{equation}  \label{fg2}
  f=\mu_+\ee^{\psi/2}\ee^{-\ii\pi t}\ee^{\ii\omega},
     \quad g=\mu_-\ee^{-\psi/2}\ee^{\ii\pi t}\ee^{-\ii\omega},
\end{equation}
where $\omega$ is some real constant.
Notice that if $\lambda_+$ and $\lambda_-$ are interchanged, then
$\psi\mapsto-\psi$, and $f$ and $g$ are interchanged. This is a gauge
transformation (\ref{gaugerefl}), so the corresponding solutions of the
Hitchin equations are gauge-equivalent. In particular, there is no
ambiguity (up to gauge-equivalence) on the cut $K\in(-2,2)$.

If $K=\pm2$, then (\ref{PsiEqn2}) admits the solution $\psi=0$, and
we get explicit solutions of the Hitchin equations for which $F_{12}$
is identically zero. Conversely, it is straightforward to show
that these are the only solutions for which $F_{12}$ is identically zero.
These solutions correspond to taking the explicit Hitchin-equation
solution for the periodic 1-monopole \cite{CK01, W05}, and recycling it
by simply doubling the period. They can also be obtained as limits of finite
monopole chains, as we show in section 5.

For $K\neq\pm2$, it is not as easy to find explicit solutions
of (\ref{PsiEqn2}). For the discussion that follows, the equation was
solved numerically, for various
values of $K$, by minimizing the functional
\begin{equation}  \label{energy1}
  E[\psi] = \int_{\Sigma}\left[\frac{1}{4}(\pa_j\psi)^2+\xi_+\ee^\psi
                  +\xi_-\ee^{-\psi} - P\right]\,dr\,dt.
\end{equation}
Here $P=P(r,t)$ is a fixed function, depending on $K$, introduced
simply in order to ensure that the integral converges. 
Note that (\ref{PsiEqn2}) is the Euler-Lagrange equation for $E[\psi]$,
so any local minimum will be a solution of (\ref{PsiEqn2}).
The functional (\ref{energy1}) was modelled using spectral methods:
Chebyshev with $n_r$ grid points for $r\in[-L,L]$,
and Fourier with $n_t$ grid points for $t$. The field $\psi$ approaches
its (finite) boundary values very rapidly --- for example,
$\psi=o(\ee^{-2\pi r})$ as $r\to\infty$ --- and there
is little loss of accuracy in taking $r$ to lie in a finite range
$[-L,L]$, with $L$ of order unity, and with $\psi$ attaining its boundary
values at $r=\pm L$.  The numerical minimization was effected
using a conjugate-gradient method, for various values of
the quantities $L$, $n_r$ and $n_t$, and the resulting fields $\psi$
compared. For any two such fields $\psi$ and $\tilde{\psi}$, we
typically obtain $||\psi-\tilde{\psi}||<0.01\,||\psi||$, where
$||\cdot||$ denotes the supremum norm $||\psi||=\max_{(r,t)}|\psi|$.
On the basis of this, we believe that for each $K$, our numerical
solution $\psi$ is within $1\%$ of the actual solution.

Given such a solution $\psi$, we then
implemented the Nahm transform numerically, to obtain the monopole fields
$(\hat{\Phi}, \hat{A}_j)$. For each point $(x,y,z)$ of a finite
3-dimensional grid, the equations (\ref{PsiEqn+}, \ref{PsiEqn-})
were solved using a relaxation method, and the monopole fields were then
computed from (\ref{MonFields}). Since $\Psi\to0$ rapidly as $r\to\pm\infty$,
it is once again reasonable to restrict $r$ to a finite range $[-L,L]$.
The partial derivative in (\ref{MonFields}) is approximated by a simple
finite difference on the grid. As a cross-check, we then sampled the
Bogomolny equations $\hat D_j\hat{\Phi}+\half\varepsilon_{jkl}\hat{F}_{kl}=0$
at various points on the grid, and specifically near the location of the
monopoles, again using finite differences for
the derivatives. For a sufficiently fine grid (both in $r,t$ and in
$x,y,z$) the accuracy is better than $1\%$, in the sense that
$|\hat D_1\hat{\Phi}+\hat{F}_{23}|<0.01\,|\hat D_1\hat{\Phi}|$ (and
analogously for the $\hat D_2\hat{\Phi}$ and $\hat D_3\hat{\Phi}$ equations).
As a further check, this time of the boundary condition, we computed
the quantity $\pi|\hat{\Phi}|/\log{x}$ on a segment of the $x$-axis;
and verified that, as required, it approaches unity as $x$ becomes large.
(In the $K=-2.3$ case, for example, the numerical result is that
$\pi|\hat{\Phi}|/\log{x}=0.986$ at $x=6.5$.) These checks
allow us to be reasonably confident that our numerical procedures
give an accurate reflection of the true solution.

The results of these numerical investigations gives the following picture.
For each $K\in\CC$, there is exactly one solution $\psi$ of (\ref{PsiEqn2});
and solutions for distinct $K$ are not gauge-equivalent. So the surface
${\cal S}$ is diffeomorphic to the plane $\CC$, on which $K$ is a global
coordinate. A solution with $|K|\gg1$ corresponds to well-separated monopoles,
located at the points $x+\ii y = \pm\sqrt{-K}$, $z=\pi$
(meaning that these are the points in $\RR^2\times S^1$ where the Higgs
field $\hat\Phi$ is zero). Each monopole is roughly spherical in shape
(see, for example, the $K=\pm4$ pictures in Figure 1).
If, at the other extreme, $K$ lies on the segment $[-2,2]$ of the real line,
then the monopoles are located on the periodic axis $x=y=0$, at
$z=\pi/4$ and $z=3\pi/4$. They are elongated in the $x$-direction
if $-2\leq K<0$ (see the $K=-2$ picture in Figure 1), and elongated in the
$y$-direction if $0<K\leq2$.

To see the transition between the large-$|K|$ and small-$|K|$ regimes,
one may examine the monopoles
corresponding to the one-parameter family $K\in\RR$. In fact, this represents
a geodesic in ${\cal S}$ (as we shall see below), and hence also a geodesic in
${\cal M}$. So this family describes
a slow-motion dynamical evolution of the system \cite{M82, MS04}, and provides
yet another example of $90^\circ$ scattering following a head-on collision ---
this time between two parallel monopole chains, scattering transversely.
See Figure 1, which plots the surface
$|\hat\Phi|^2 := \tr(\hat\Phi \,\hat\Phi^*) = 0.004$,
for the monopole solutions corresponding to various real values of $K$.
\begin{figure}[htb]
\begin{center}
\includegraphics[scale=1.0]{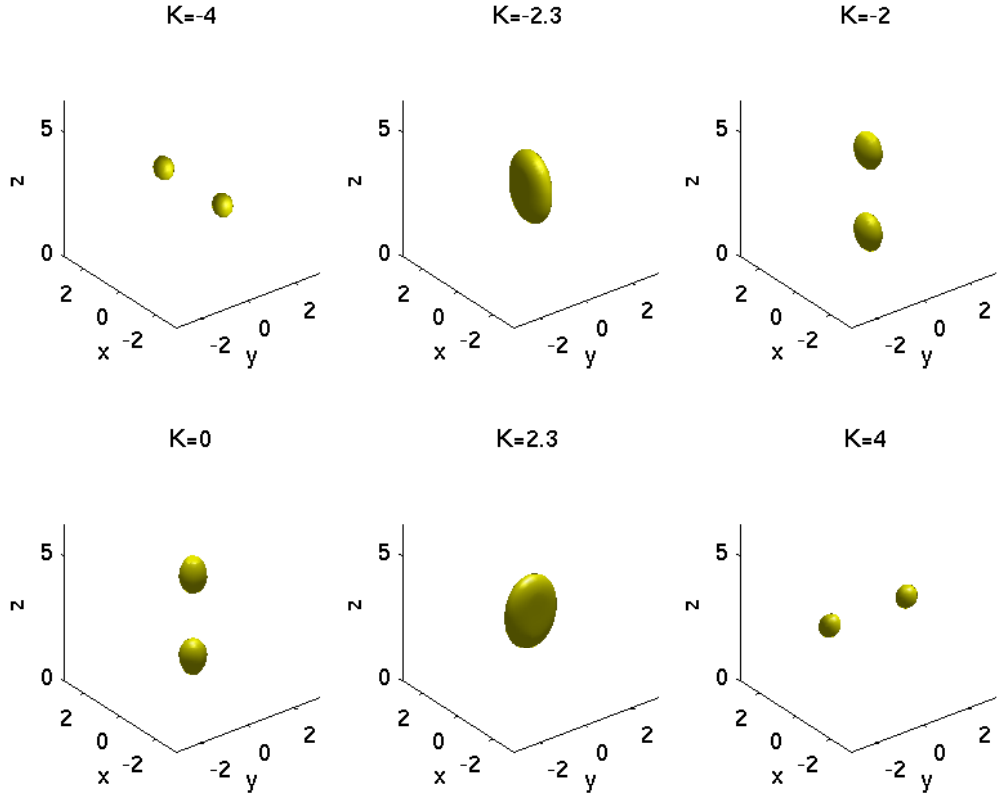}
\caption{two-monopole scattering --- snapshots for $K=-4, -2.3, -2, 0, 2.3, 4$.
 \label{fig1}}
\end{center}
\end{figure}
The interpretation in terms of parallel monopole chains, is that the chains
approach each other in the $x$-direction; the individual monopoles
coalesce and then separate to form a single chain of 1-monopoles; and
finally the monopoles in this chain re-coalesce and then separate to
form two parallel chains which move apart in the $y$-direction.

To see that $K\in\RR$ is a geodesic, we show that the map $K\mapsto\overline{K}$
is an isometry. First, note that the isometry of $\RR^2\times S^1$ given by
$x\mapsto x$, $y\mapsto-y$, $z\mapsto2\pi-z$ preserves the field equation
and boundary condition for periodic monopoles. So this map induces an isometry
on $\widehat{\cal M}$. The corresponding map in the
Hitchin-equation space is $\Phi(r,t)\mapsto\Phi(r,1-t)^*$, 
$A_r(r,t)\mapsto A_r(r,1-t)$, $A_t(r,t)\mapsto-A_t(r,1-t)$,
  $K\mapsto\overline{K}$,
which preserves the Hitchin equations and its associated constraints.
The corresponding map on $\Psi_{\pm}$ is
$\Psi_{\pm}(r,t;x,y,z)=\Psi_{\mp}(r,1-t;x,-y,2\pi-z)$. So $K\mapsto\overline{K}$
is an isometry of ${\cal S}\subset{\cal M}$, and its fixed set $\Im{(K)}=0$ is
a geodesic. These monopole solutions are invariant under the
Klein group $D_2$, consisting of rotations by $\pi$ about each of the
three coordinate axes. The solution for $K=0$ has additional symmetry,
namely rotation by $\pi/2$ about the $z$-axis.

In the asymptotic region $|K|\gg1$, the metric on the moduli space is
explicit, and simple to write down \cite{CK02}; but in the interior
region, it is likely to be rather complicated. As a
starting-point, we have seen that there is a totally-geodesic surface
${\cal S}$ in the moduli space, representing to periodic 2-monopoles
which are invariant under rotation by $\pi$ about the periodic axis;
and that these correspond, via the Nahm transform,
to real-valued functions $\psi(r,t)$ satisfying (\ref{PsiEqn2}), together
with appropriate boundary conditions. One geodesic in ${\cal S}$ represents
a head-on collision of two monopole chains. It should be feasible to extend
the procedure used in this note, to learn more about the metric and its
geodesics, and hence about more general scattering processes.

%%%%%%%%%%%%%%%%%%%%%%%%%%%%%%%%%%%%%%%%%%%%%%%%%%%%%%%%%%%%%%%%%%%%%

\section{Limits of Finite Chains}

In this section, we will examine how the infinite chain of 1-monopoles
arises as the $N\to\infty$ limit of a finite chain of N single monopoles.
One knows that this limit is delicate, and that care is needed to avoid
divergences ({\sl cf.}\ \cite{L99, CK01, DK05}). The aim here is to describe
the limit in the Nahm-transformed picture.
The Nahm data for the $N$-chain are known \cite{ES89}, and we will see how
to take the $N\to\infty$ limit, thereby obtaining the Hitchin data
corresponding to an infinite chain. In fact, one may regard this monopole
field as being a chain of $m$-monopoles, for any integer $m\geq1$,
with U($m$)-valued Hitchin data; we shall do the two cases $m=1$
and $m=2$, and the latter therefore links up with our earlier material on
2-monopole chains.

The Nahm data for a monopole of charge $N$ consists of three anti-Hermitian
matrix-valued functions $T_a(u)$, for $a=1,2,3$ and $u\in (-1,1)$, which
satisfy the Nahm equation
$dT_a/du = \half\epsilon_{abc} [T_b,T_c]$, plus appropriate boundary
conditions. The Nahm data for an $N$-chain have the form
$T_a(u) = -if_a(u) J_a$, where $J_a$ are the generators of the $N$-dimensional
irreducible representation of SU(2), and satisfy
$[J_a,J_b]=i\epsilon_{abc}J_c$. Explicitly, we may set $J_1=\half(J_- + J_+)$,
$J_2=\half\ii(J_--J_+)$ and $(J_3)_{ij}=-j\delta_{i,j}$, where
\begin{eqnarray*}
  (J_+)_{ij} &=& \half\delta_{i+1,j}\sqrt{(N+1+2i)(N+1-2j)}, \\
  (J_-)_{ij} &=& \half\delta_{i,j+1}\sqrt{(N+1+2j)(N+1-2i)}.
\end{eqnarray*}
The indices $i,j$ run from $-(N-1)/2$ to $(N-1)/2$, and take values in
$\mathbb{Z}$ or $\mathbb{Z}+1/2$ according to whether $N$ is odd or even.
The functions $f_a$ are specified in terms of Jacobi elliptic functions by
\[
  f_1(u)=K\,k'\,nc(Ku;k), \quad f_2(u)=K\,k'\,sc(Ku;k), \quad
    f_3(u)=K\, dc(Ku;k),
\]
where $k$ is the elliptic modulus, $k'=\sqrt{1-k^2}$ the complementary
modulus, and $K(k)$ the complete elliptic integral of the first kind.
These Nahm data correspond to a finite-length collinear chain of $N$
single monopoles,
and the parameter $k$ determines the ratio between the monopole size
and the distance between adjacent monopoles in the chain.

Now define a new variable $r=Ku/(2\pi)$. Then $T'_a = 2\pi T_a/K$ solve
the rescaled Nahm equation
\begin{equation}\label{Nahm equation}
 \frac{d}{dr}T_a' = \half\epsilon_{abc} [T_b',T_c'].
\end{equation}
We restrict attention to the cases where $N$ is odd, and consider the
limit $N\rightarrow\infty$, $k'\rightarrow 0$, with
$\pi Nk'\rightarrow L$ for some constant $L\in\mathbb{R}$. In this limit
the matrices $T_a'$ become infinite-dimensional, and naturally operate on
functions $u(r,t)$, periodic in $t$, with Fourier expansion
$u = \sum_{j\in\mathbb{Z}} u_j(r) \exp(2\pi\ii jt)$. Note that the range of
$r$ becomes $(-\infty,\infty)$ in the limit. Using
known limits of elliptic functions, we obtain
\begin{eqnarray*}
  \left( \lim T_1' \right) \cdot u &=& -A_4(r,t)u(r,t), \\
  \left( \lim T_2' \right) \cdot u &=& A_3(r,t)u(r,t), \\
  \left( \lim T_3' \right) \cdot u &=& \partial_t u(r,t),
\end{eqnarray*}
where $A_3(r,t) = \ii L\sinh(2\pi r)\sin(2\pi t)$ and
$A_4(r,t) = \ii L\cosh(2\pi r)\cos(2\pi t)$ . The limit of the Nahm equation
(\ref{Nahm equation}) can be rewritten as the operator equation
\begin{equation}
[ \partial_r, \lim T_a' ] = \half\epsilon_{abc} [ \lim T_b', \lim T_c' ],
\end{equation}
which is equivalent to the Hitchin equations (\ref{H1}, \ref{H2}) under the
identifications
$A_1=0$, $\partial_t + A_2= \lim T_3'$, and $\Phi = A_3-\ii A_4$. Explicitly, we
obtain the known solution of the Hitchin equations corresponding to a
charge 1 periodic monopole, namely $A_1=0$, $A_2=0$, $\Phi=L\cosh(2\pi s)$.

As mentioned above, one may vary this procedure to obtain Hitchin data
for an infinite chain of $m$-monopoles, starting from the same Nahm data;
here is the $m=2$ version. This time, define $r=Ku/\pi$, so that
$T_a'= \pi T_a/K$ solve (\ref{Nahm equation}). Restrict attention to the
cases where $N$ is even, and consider again the limit where
$N\rightarrow\infty$, $k'\rightarrow 0$, with $\pi Nk'\rightarrow L$ constant.
The resulting infinite matrices act naturally on vectors $u_j$ with
$j\in\mathbb{Z}+\frac{1}{2}$, and hence on functions
\[
  u(r,t) = \sum_{k\in\mathbb{Z}} \left( \begin{array}{c}
    u_{2k-\frac{1}{2}}(r) \\
    u_{2k+\frac{1}{2}}(r)
     \end{array} \right) \exp(2\pi \ii k t).
\]
A direct calculation yields
\begin{eqnarray*}
\left( \lim T_1' \right) \cdot u &=& -A_4(r,t)u(r,t), \\
\left( \lim T_2' \right) \cdot u &=& A_3(r,t)u(r,t), \\
\left( \lim T_3' \right) \cdot u &=& \left( \frac{\partial}{\partial t} - \frac{\ii\pi}{2} \left( \begin{array}{cc} 1 & 0 \\ 0 & -1 \end{array} \right) \right) u(r,t), \\
A_4(r,t) &:=& \frac{\ii L}{2} \cosh(\pi r)\cos(\pi t) \left( \begin{array}{cc} 0 & e^{\pi \ii t} \\ e^{-\pi \ii t} & 0 \end{array} \right), \\
A_3(r,t) &:=& \frac{\ii L}{2} \sinh(\pi r)\sin(\pi t) \left( \begin{array}{cc} 0 & e^{\pi \ii t} \\ e^{-\pi \ii t} & 0 \end{array} \right).
\end{eqnarray*}
Hence we obtain Hitchin data
\[
  A_1=0, \quad
  A_2=-\frac{\ii\pi}{2}\left(\begin{array}{cc}
          1 & 0 \\ 0 & -1 \end{array} \right), \quad
  \Phi=\frac{L}{2}\cosh(\pi s)\left(\begin{array}{cc}
          0 & e^{\pi \ii t} \\ e^{-\pi \ii t} & 0 \end{array} \right).
\]
With $L=4$, these are gauge-equivalent to the explicit solution
$\alpha=0$, $K=-2$, $\psi=0$ mentioned in section 4. The above discussion
justifies the claim that, in this case, the 2-monopole chain is just
a 1-monopole chain in disguise.

\subsubsection*{Acknowledgement}
Derek Harland acknowledges the support of an STFC studentship while some of this work was completed.

%%%%%%%%%%%%%%%%%%%%%%%%%%%%%%%%%%%%%%%%%%%%%%%%%%%%%%%%%%%%%%%%%%%%

\end{document}